# Beyond ensemble averaging: Parallelized single-shot readout of hole capture in diamond

Richard Monge[1], Yuki Nakamura[3], Olaf Bach[1], Jason Shao[1], Artur Lozovoi[1,*], Alexander A. Wood[4], Kento Sasaki[3], Kensuke Kobayashi[3], Tom Delord[1,†], and Carlos A. Meriles[1,2,†]

Understanding the generation, transport and capture of charge carriers in semiconductors is of fundamental technological importance. However, the ensemble measurement techniques ubiquitous in electronics offer limited insight into the nanoscale environment that is crucial to the operation of modern quantum-electronic devices. Here, we combine widefield optical microscopy with precision spectroscopy to examine the capture of photogenerated holes by negatively charged nitrogen vacancy (NV⁻) centers in diamond. Simultaneous single-shot charge readout over hundreds of individual NVs allows us to resolve the roles of ionized impurities, reveal the formation of space charges fields, and monitor the thermalization of hot photo-carriers during diffusion. We measure effective NV⁻ hole capture radii in excess of 0.2 µm, a value approaching the Onsager limit and made possible here thanks to the near-complete neutralization of coexisting charge traps. These results establish a new platform for resolving charge dynamics beyond ensemble averages, with direct relevance to nanoscale electronics and quantum devices.

Charge diffusion and capture in semiconductors is central to modern technology, yet our ability to resolve these dynamics at the nanoscale remains experimentally challenging. Traditional measurements rely on macroscopic observables, which average over many events and mask the local physics governing device performance. This gap is increasingly consequential in quantum technologies, where local charge instabilities and field noise directly limit the coherence times and fidelity of solid-state qubits and quantum sensors[1-3]. At the same time, charge-based interactions offer new functionality: Mobile carriers can mediate long-range coupling[4], enhance quantum sensitivity[5], or act as buses connecting otherwise isolated quantum nodes[6-11]. Gaining direct experimental access to these charge processes at the single-particle level is thus essential, not only to refine models of transport and recombination, but to realize robust, scalable quantum systems.

Quantum sensors now provide powerful tools to meet this challenge[12]. Among the most versatile are optically addressable color centers such as the nitrogen-vacancy (NV) center in diamond[13], which offers atomic-scale spatial resolution combined with exquisite sensitivity to magnetic fields[14], temperature[15], and local electric environments[16,17]. These capabilities have enabled applications ranging from nanoscale magnetometry[18] to biological thermometry[19]. More recently, they have been harnessed to probe semiconductor charge dynamics, locating individual charge traps[20-22], detecting local space-charge formation[23,24], and observing hole transport under ambient conditions[25,26].

In this article, we utilize a collection of individually addressable color center sensors to measure the diffusion and capture of single photo-injected holes; we work under cryogenic conditions, which enables a regime of long-range transport and enhanced sensitivity. Specifically, we exploit resonant excitation of color centers to simultaneously track the charge state and electric environment of hundreds of charged defects individually and non-destructively. This platform reveals singular features of low temperature carrier dynamics: With phonon scattering nearly suppressed, carrier diffusion becomes dominated by ionized impurities and space charge fields, which we directly observe via NV electrometry. Further, we demonstrate controlled neutralization of these charged traps, improving carrier mobility, and minimizing electric noise; in this limit we demonstrate hole capture radii as large as ~0.2 µm, approaching the Onsager radius. Lastly, we find that carrier capture rates increase with the distance traveled by carriers, which is consistent with the photo-injection of hot holes and their slow thermalization over nanoseconds[27-29].

### Single-shot readout of NV⁻ hole capture

Our experiments combine resonant excitation of NV centers at low temperature (9K) with widefield microscopy[30]. Leveraging the high-quality factor of NV optical resonances (~5×10⁷) this approach enables non-destructive, multiplexed, single shot readout of NV charge states over wide areas. Figure 1a illustrates our measurement protocol: Following widefield green (532 nm) illumination to charge initialize NVs into the negatively charged state, we apply short pulses of focused green light to charge cycle a "source" NV from negative to neutral and back, in the process injecting a stream of electrons and holes[31]. To track the propagation of these carriers, we interlace steps of charge state readout with the help of a charge-coupled device (CCD) camera and a narrowband 637 nm laser swept in wavelength[32] (see Supplementary Material (SM), Section I). Figure 1b shows the resulting photoluminescence (PL) from $NV_S$, the source NV, as a function of the integrated green illumination time: We observe cycles through negative ($NV^-$) and neutral ($NV^0$) states — respectively bright and dark under 637 nm readout — revealing events of photoionization and recombination under green illumination[31]. The two states are cleanly separated in photon count space, enabling charge-state discrimination with high fidelity (>99.9%; Fig. 1c).

[1]Department of Physics, CUNY- The City College of New York, New York, NY 10031, USA. [2]CUNY-Graduate Center, New York, NY 10016, USA. [3]Department of Physics, The University of Tokyo, Bunkyo-ku, Tokyo, 113-0033, Japan. [4]School of Physics, University of Melbourne, Parkville, Victoria 3010, Australia. *Present address: Department of Electrical and Computer Engineering, Princeton University, Princeton, New Jersey 08544, USA. †E-mails: tdelord@ccny.cuny.edu, cmeriles@ccny.cuny.edu.

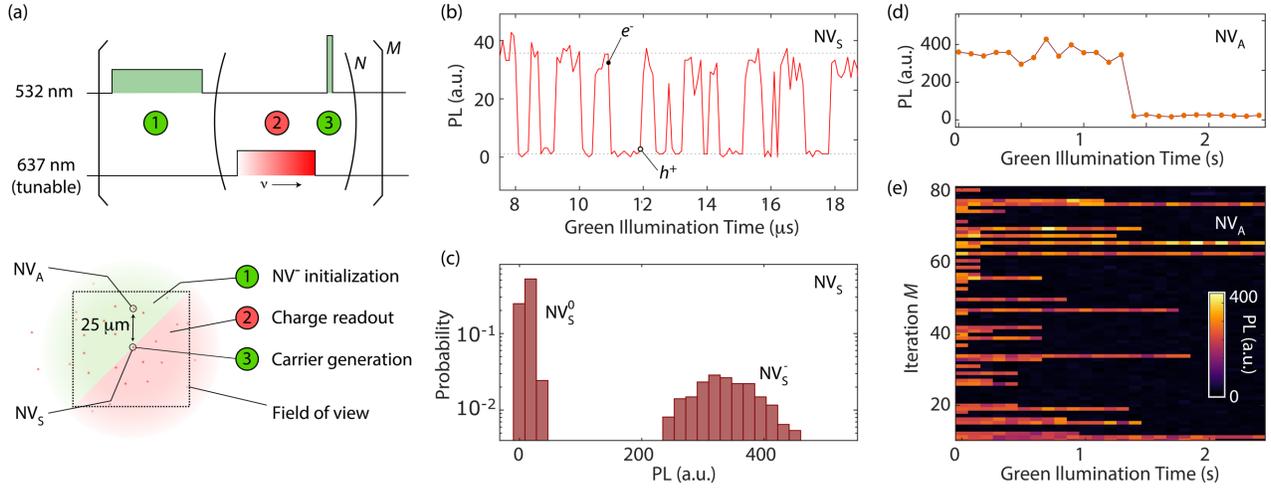

**Figure 1 | Detection of individual hole capture events.** (a) Measurement protocol; upon NV charge initialization via widefield green excitation (①), we recurrently alternate weak red illumination of tunable frequency $\nu$ (②) and local green excitation (③) to respectively record the spectra from all NVs within the field of view and cycle the charge state of a source color center, $NV_S$. (b) PL trace from $NV_S$ as a function of the green illumination time; a transition from bright to dark (dark to bright) heralds the emission of an electron, $e^-$ (respectively, a hole, $h^+$). (c) Probability distribution of photon emission from $NV_S$ as derived from (b); we attain a charge state readout fidelity exceeding 99.9%. (d) PL brightness of an NV ~25 μm away from $NV_S$ (labeled $NV_A$ in the schematics in (a)) as a function of the integrated green illumination time. Sudden transition to a dark state signals the capture of an itinerant hole. (e) Multiple repetitions of the experiment in (c).

Figure 1d extends the same protocol to $NV_A$, another color center in the set approximately 25 μm away from $NV_S$. Starting from the bright (negatively-charged) state, the PL trace displays a sudden drop in brightness consistent with the capture of a single hole and subsequent transition to the neutral $NV^0$ state. Notice this process is irreversible because, in the absence of Coulombic attraction, the electron capture probability by $NV^0$ is negligible. Crucially, repeated single-shot readout of the same individual NV reveals the stochastic nature of the hole capture process (Fig. 1e); these experiments, therefore, expose the richer physics at play, only partly probed in conventional transport experiments that average over many events and obscure microscopic variability.

We can build on these measurements in various ways, for example, to produce time distribution functions and extract the hole capture probability for all individual NVs in the set. Figure 2a starts with representative results for $NV_A$ as well as $NV_B$, a second color center 14 μm away from the carrier source $NV_S$. In both cases, the capture probability distribution appears exponential because we only chose to resolve the long-time tail of the underlying Poisson process. We fit this tail to extract a characteristic capture time $\tau$ that varies with distance.

Generalizing this approach to the entire ensemble, we construct a spatial map of capture rates $\omega \equiv 1/\tau$ (Fig. 2b), revealing a dependence on proximity to the charge source. Note the cryogenic conditions in use are particularly beneficial for these experiments because the mean free path can reach tens of microns[29]. We find higher capture rates for NVs closer to the carrier source, consistent with the carrier flux decaying with distance. While our analysis focused on NVs in the same focal plane as the source, the method readily extends to other planes, enabling full three-dimensional mapping[26]. Local fluctuations aside, their dependence reveals a $1/d^2$ decay, characteristic of lossless isotropic hole propagation (Fig. 2c).

To quantify the trapping efficiency of individual defects, we introduce the hole capture cross section[33,34] $\sigma_h = \pi r_h^2$, expressed in terms of the capture radius $r_h$. These parameters relate to the measured capture rate via the relationship $\omega = k_h \sigma_h / (4\pi d^2)$, where $k_h$ is the hole generation rate at the source. The extracted capture radii

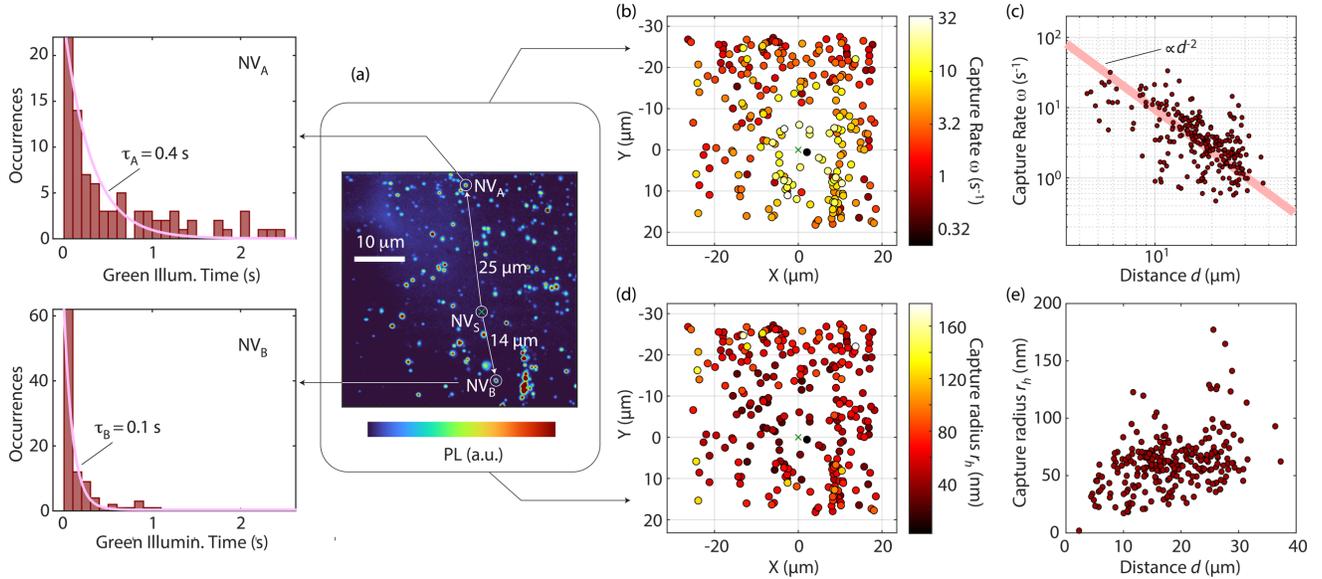

**Figure 2 | Statistics of hole capture.** (a) Hole capture probability distributions as a function of the green illumination time for $NV_A$ and $NV_B$, respectively at 25 and 14 μm from the charge source, $NV_S$; from exponential fits (pink traces), we extract characteristic hole capture times $\tau_A \cong 0.4$ s and $\tau_B \cong 0.1$ s. (b) Capture rate for the NVs within the field of view; centers proximal to $NV_S$ (signaled with a green cross) tend to exhibit higher capture rates. (c) Capture rate as a function of distance $d$ to $NV_S$ as extracted from the capture rate map in (c). For reference, the red line represents an inverse square dependence. (d) Effective hole capture radius for the NVs in (b). (e) Same as in (d) but as a function of the distance to $NV_S$.

— displayed in Figs. 2d and 2e — show a spread, ranging from ~20 to ~120 nm. Interestingly, we observe an increasing trend with distance to the source, which we leverage to examine hole capture in NVs up to 37 μm away. Combined with the high spatial density of sensing NVs, these large capture cross-sections enable the detection of approximately one hole out of every 3000 generated.

The capture radius $r_h$ is expected to increase as the carrier thermal kinetic energy diminishes[35], growing from ~30 nm at room temperature[25] to nearly 300 nm at 9K. Yet our measurements reveal a more nuanced picture: The extracted $r_h$ values fall short of this expectation and vary significantly across different NVs. This spatial inhomogeneity points to a strong influence of the local electrostatic environment and defect configuration, an aspect we explore in more detail below.

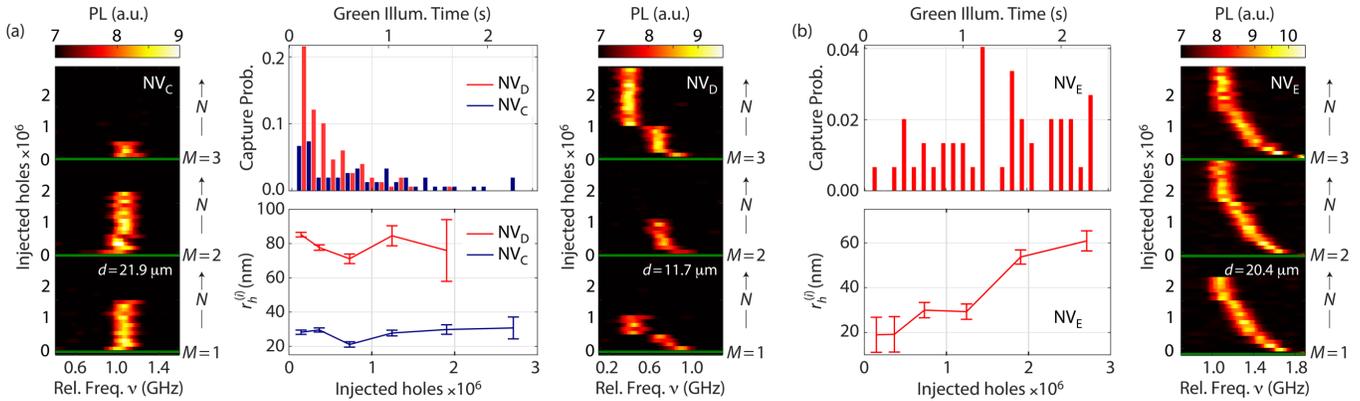

**Figure 3 | Detection of space charge fields.** (a) PLE spectra, hole capture probability histograms, and "instantaneous" hole capture radius $r_h^{(i)}$ for two NVs in the set. While the $NV_C$ spectrum — shown here for measurement sets following green charge initialization (horizontal green lines, see $M$ and $N$ notation in Fig. 1a) — remains relatively stable over time, the equivalent optical resonance of $NV_D$ undergoes a rapid shift followed by a discrete jump. Nonetheless, the capture probability distribution of both NVs shows exponential behavior, leading to a stable instantaneous capture radius $r_h^{(i)}$ throughout the green illumination time (here converted to the number of injected holes). (b) Same as in (a) but for a third color center, $NV_E$. Unlike the case in (a), $r_h^{(i)}$ grows for longer green illumination times, while the PLE spectrum shows a gradual shift to lower frequencies, consistent with progressive charge neutralization of the local environment. The reference frequency in the spectra of (a) and (b) is 470.474 THz.

**Space-charge field formation and trap neutralization**

Given the long distances separating the source from the probe NVs and large capture radii, even low concentrations of background impurities should influence the hole capture dynamics, suggesting one could leverage the present approach to monitor changes in the local electrostatic environment. Figures 3a and 3b compare the photoluminescence excitation (PLE) spectra and capture probabilities for two NVs, C and D, showing markedly different behavior. Specifically, while $NV_C$ remains spectrally stable during repeated hole injection, $NV_D$ exhibits both discrete and continuous redshifts, signaling local electric field changes via the Stark effect. These shifts — indicative of local electric field changes — reveal the presence of nearby space-charge regions evolving as carriers are trapped. In particular, the spectral changes in $NV_D$ suggest the presence of nearby charged defects — likely substitutional nitrogen or boron — within a few hundred nanometers, which can either screen the $NV^-$ Coulomb potential or compete directly for holes, thereby suppressing capture. Such interactions with proximal charge traps provide a plausible explanation for the variability in $\sigma_h$ observed between otherwise similar NV centers.

This relationship between the local charge environment and capture efficiency holds broadly across the sample. NVs with smaller capture cross-sections are eight times more likely to exhibit systematic spectral shifts compared to NVs with above-average capture radii (40% vs 5%). In some cases, this correlation can be observed within a single NV as the local environment evolves. For example, Fig. 3b shows an instance where the NV experiences a gradual spectral shift over time. Here, the hole capture probability distribution deviates from simple exponential behavior; instead, the "instantaneous" capture radius $r_h^{(i)}$ — here extracted from the fractional number of capture events at a given time, see SM, Section II — increases linearly, coinciding with the progressive neutralization of nearby space charge captured by the time-dependent PLE spectrum.

Based on the assumption that a sustained flux of electrons and holes from the source NV fills all electron and hole traps in the vicinity of the probe NV[36], we now exploit these observations to purposely engineer the charge environment: We modify the $NV^-$ initialization protocol to include a neutralization step, based on carrier injection from a remote NV source followed by local charge initialization of the probe NV of choice; the latter relies on a focused 2 μs green pulse sufficiently short to minimally disrupt the neutralized environment (Figs. 4a and 4b). Recurrent optical spectroscopy reveals a nearly lifetime-limited inhomogeneous linewidth of 28 MHz, with a five-fold reduction in the resonance frequency standard deviation (from 36 to 7 MHz). This result is consistent with Monte Carlo simulations of the charge environment (SM, Section III), indicating a ~25-fold reduction in charge density, from approximately 6 to 0.15 electrons per cubic micron. Crucially, this environmental control also affects capture dynamics: The widefield protocol of Fig. 4c yields a slow, exponential decay in hole capture probability, with a time constant of $\tau = 4.96$ s, consistent with hindered transport in a disordered charge landscape. In contrast, the protocol in Fig. 4d yields a much shorter capture time, $\tau = 0.31$ s, indicating enhanced hole mobility in a neutralized environment.

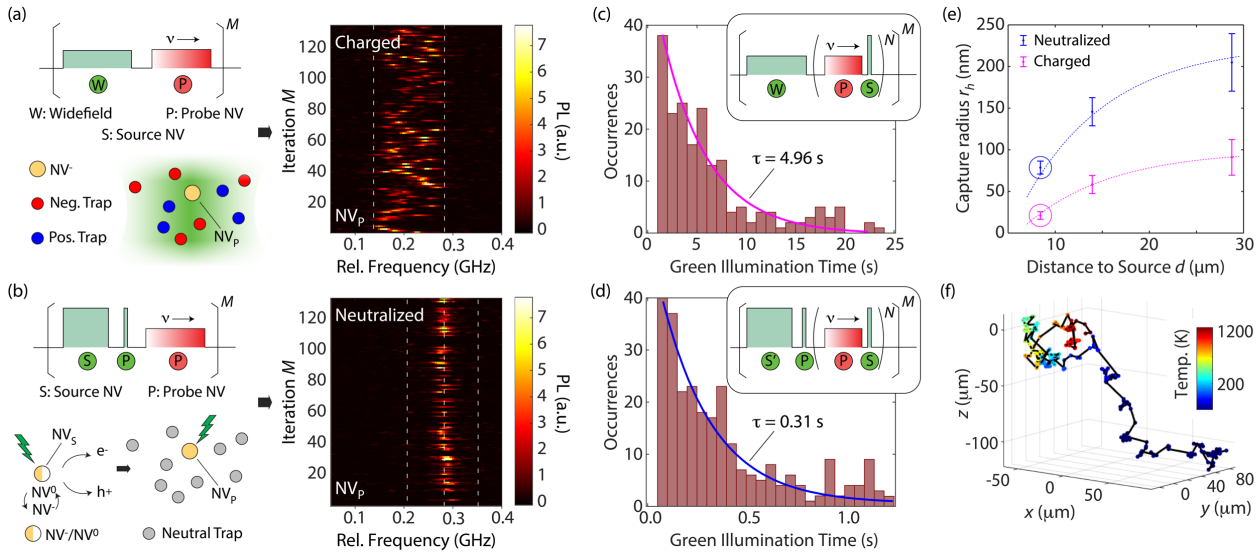

**Figure 4 | Controlling the electrical environment.** (a) Recurrent optical spectroscopy of a probe color center, $NV_P$, after widefield charge initialization. Ionization of proximal traps under green light leads to significant spectral diffusion. Here (and everywhere below), green (red) blocks indicate 532 mn (637 nm) laser illumination; white dashed lines are guides to the eye. (b) Same as in (a) but for a different charge initialization protocol comprising focused green illumination of a charge source, $NV_S$, followed by local charge initialization of $NV_P$; neighboring traps neutralize upon capture of photo-injected carriers leading to much reduced spectral diffusion of the probe NV. (c) NV hole capture probability distribution using the charge initialization protocol in (a); we measure a capture time $\tau = 4.96$ s. (d) Same as in (c) but using the charge initialization protocol in (b); the reduced capture time, $\tau = 0.31$ s, suggests improved hole mobility. Note that charge injection during initialization and hole capture measurement, respectively denoted by steps S′ and S in the schematic, do not necessarily rely on the same NV (SM, Section II). (e) Effective hole capture radius as derived from the protocols in (c) and (d) (magenta and blue, respectively) for NVs at variable distances from the source. Magenta and blue circles respectively denote the values extracted in (c) and (d), and dashed lines are guides to the eye. (f) Monte Carlo simulation of hole thermalization in diamond; carriers diffuse tens of microns before they cool down to reach the lattice temperature.

Figure 4e compares the impact of environment neutralization on carrier transport for three probe NVs at different distances from the source. Across all measured NVs, the neutralization protocol yields a nearly tenfold increase in hole capture radius compared to the charged case, highlighting the strong influence of electrostatic disorder on carrier capture. In both protocols, however, the effective capture radius increases with distance from the source NV (Fig. 4e), indicating a shared underlying mechanism. That this trend persists even in the neutralized case — where disorder is largely suppressed — suggests additional limits to capture efficiency near the source.

We explain and model this surprising behavior by considering the excess energy of photo-injected holes and the suppressed relaxation rates at low temperatures. As depicted in Fig. 4e, hot carriers are injected at the source NV[37,38] and thermalize through random phonon scattering as they diffuse over tens of microns[27,28]. Our model captures the effect of both carrier thermalization and the inhomogeneity arising from screening by (or competition with) coexisting traps (SM, Section III); the lower capture cross-section at short distances reflect the carrier initial kinetic energy, consistent with the laser frequency and ab initio calculations of the ionization threshold of the source defect[37]. Interestingly, the ~200 nm capture radius we attain under optimal conditions approaches the Onsager value at 9 K (~300 nm), reflecting the possible formation of transient Rydberg states[25,39,40] and of interest as a route to carrier transfer between proximal NVs[20]. We note this strategy is not diamond-specific; analogous color centers in materials like silicon[41,42] or SiC[43] offer immediate pathways to study carrier dynamics in other technologically relevant semiconductors.

## Conclusions and outlook

Our results establish a new approach for studying carrier dynamics in semiconductors by leveraging quantum sensors as nanoscale probes. Using resonant spectroscopy of NV centers in cryogenic diamond, we directly observe individual hole capture events, in the process revealing how electric disorder and carrier thermalization govern their fate. We find that electrostatic fields from charged background defects can strongly suppress capture, even in nominally pure materials, and that neutralizing these traps dramatically enhances both mobility and NV capture probability, boosting the effective capture rate by an order of magnitude. We further uncover a surprising distance dependence, with capture efficiency increasing as holes diffuse and cool down away from their injection site. By selecting thermalized carriers and through careful control of the charge environment, we demonstrated record capture cross sections (0.13 μm$^2$) and stabilized optical resonances within one lifetime-limited linewidth, crucial to quantum networking with color centers[44].

These findings highlight the utility of quantum sensors as active diagnostic tools, capable not just of detecting local fields, but of interrogating and controllably altering the microscopic mechanisms underlying charge flow in heterogeneous media. A key feature of our approach is the use of repeated single-shot measurements, which enables access to rare events and heterogeneity. Looking ahead, this platform provides a powerful framework for studying correlated charge dynamics across large color center sets[20]; these correlations could be exploited to reveal collective behavior, spatial transport anisotropies, or transient charge clustering effects, opening the door to statistical analyses that are otherwise inaccessible. Tailored device architectures and external gating could enable mapping of mobility in engineered potential landscapes[26], while integration with spin-based sensing may allow simultaneous tracking of charge and spin degrees of freedom.

More broadly, this work paves the way toward active control of charge environments in quantum devices[45,46], and toward hybrid platforms where mobile carriers are harnessed as dynamic links between solid-state quantum systems[6-11,47], e.g., via cold carriers injected via near-threshold photo-emission[48] or coherent transfer[49]. Combined with time-resolved excitation, device integration, or hybrid sensing modalities, this capability could enable new studies of charge flow in complex or disordered media with unprecedented detail. One area of particular technological importance is the investigation of hot carrier thermalization[50] and ionized impurity scattering, of fundamental interest for innovation in photovoltaic cells[51-53].

## Data availability

The data that support the findings of this study are available from the corresponding author upon reasonable request.

**Code availability**

All source codes for data analysis and numerical modeling used in this study are available from the corresponding author upon reasonable request.

**Author contributions**

T.D., R.M. and C.A.M. designed the experimental platform and protocols. R.M. and T.D. constructed the apparatus and performed the experiments. T.D., Y.N. and J.S. performed the data analysis. Y.N., O.B. and T.D. performed the simulations. A.A.W. and A.L. provided support interpreting the data. T.D. and C.A.M. wrote the manuscript with input from all authors. C.A.M. supervised the research.

**Acknowledgments**

T.D. acknowledges support by the U.S. Department of Energy, Office of Science, National Quantum Information Science Research Centers, Co-design Center for Quantum Advantage (C2QA) under contract number DE-SC0012704. A.L. and C.A.M. acknowledge support from the National Science Foundation (NSF) via grant NSF-2216838. O.B. and J.S. acknowledge support from NSF via grant NSF-2328993. R.M. acknowledges support from NSF via grant NSF-2316693. K.S., and K.K. acknowledge partial support by JST CREST Grant Number JPMJCR23I2, Japan, and Y.N. acknowledges financial support from FoPM, the WINGS Program, The University of Tokyo. K.S. also acknowledges support from Grants-in-Aid for Scientific Research No. JP25K00934. A.A.W. acknowledges support from the Australian Research Council (DE210101093). All authors acknowledge access to the facilities and research infrastructure of the NSF CREST IDEALS, grant number NSF-2112550.

**Competing interests**

The authors declare no competing interests.

**Correspondence**

Correspondence and requests for materials should be addressed to T.D. or C.A.M.
**References**

[1] D. Bluvstein, Z. Zhang, A.C.B. Jayich, "Identifying and mitigating charge instabilities in shallow diamond nitrogen-vacancy centers", *Phys. Rev. Lett.* **122**, 076101 (2019).

[2] L. Orphal-Kobin, K. Unterguggenberger, T. Pregnolato, N. Kemf, M. Matalla, R.-S. Unger, I. Ostermay, G. Pieplow, T. Schröder, "Optically coherent nitrogen-vacancy defect centers in diamond nanostructures", *Phys. Rev. X* **13**, 011042 (2023).

[3] B.A. Myers, A. Ariyaratne, A.C.B. Jayich, Double-quantum spin-relaxation limits to coherence of near-surface nitrogen-vacancy centers", *Phys. Rev. Lett.* **118**, 197201 (2017).

[4] Y. Fang, P. Philippopoulos, D. Culcer, W.A. Coish, S. Chesi, "Recent advances in hole-spin qubits", *Mater. Quantum Technol.* **3**, 012003 (2023).

[5] A. Lozovoi, G. Vizkelethy, E. Bielejec, C.A. Meriles, "Imaging dark charge emitters in diamond via carrier-to-photon conversion", *Sci. Adv.* **8**, eabl9402 (2022).

[6] F. van Riggelen-Doelman, C.-A. Wang, S.L. de Snoo, W.I.L. Lawrie, N.W. Hendrickx, M. Rimbach-Russ, A. Sammak, G. Scappucci, C. Déprez, M. Veldhorst, "Coherent spin qubit shuttling through germanium quantum dots", *Nat. Commun.* **15**, 5716 (2024)

[7] T.A. Baart, M. Shafiei, T. Fujita, C. Reichl, W. Wegscheider, L.M.K. Vandersypen, "Single-spin CCD", *Nat. Nanotechnol.* **11**, 330 (2016).

[8] H. Flentje, P.-A. Mortemousque, R. Thalineau, A. Ludwig, A.D. Wieck, C. Bäuerle, T. Meunier, "Coherent long-distance displacement of individual electron spins", *Nat. Commun.* **8**, 501 (2017).

[9] A.R. Mills, D.M. Zajac, M.J. Gullans, F.J. Schupp, T.M. Hazard, J.R. Petta, "Shuttling a single charge across a one-dimensional array of silicon quantum dots", *Nat. Commun.* **10**, 1063 (2019).

[10] J. Yoneda, W. Huang, M. Feng, C.H. Yang, K.W. Chan, T. Tanttu, W. Gilbert, R.C.C. Leon, F.E. Hudson, K.M. Itoh, A. Morello, S.D. Bartlett, A. Laucht, A. Saraiva, A.S. Dzurak, "Coherent spin qubit transport in silicon", *Nat. Commun.* **12**, 4114 (2021).

[11] A.M.J. Zwerver, S.V. Amitonov, S.L. de Snoo, M.T. Mądzik, M. Rimbach-Russ, A. Sammak, G. Scappucci, L.M.K. Vandersypen, "Shuttling an electron spin through a silicon quantum dot array", *PRX Quantum* **4**, 030303 (2023).

[12] C.L. Degen, F. Reinhard, P. Cappellaro, "Quantum sensing", *Rev. Mod. Phys.* **89**, 035002 (2017).

[13] M.W. Doherty, N.B. Manson, P. Delaney, L.C.L. Hollenberg, "The negatively charged nitrogen-vacancy centre in diamond: the electronic solution", *New J. Phys.* **13**, 025019 (2011).

[14] G. Balasubramanian, I.Y. Chan, R. Kolesov, M. Al-Hmoud, J. Tisler, C. Shin, C. Kim, A. Wojcik, P.R. Hemmer, A. Krueger, T. Hanke, A. Leitenstorfer, R. Bratschitsch, F. Jelezko, J. Wrachtrup, "Nanoscale imaging magnetometry with diamond spins under ambient conditions", *Nature* **455**, 648 (2008).

[15] V.M. Acosta, E. Bauch, M.P. Ledbetter, C. Santori, K.-M.C. Fu, P.E. Barclay, R.G. Beausoleil, H. Linget, J.F. Roch, F. Treussart, S. Chemerisov, W. Gawlik, D. Budker, "Temperature dependence of the nitrogen-vacancy magnetic resonance in diamond", *Phys. Rev. Lett.* **104**, 070801 (2010).

[16] Ph. Tamarat, T. Gaebel, J.R. Rabeau, M. Khan, A.D. Greentree, H. Wilson, L.C.L. Hollenberg, S. Prawer, P. Hemmer, F. Jelezko, J. Wrachtrup, "Stark shift control of single optical centers in diamond", *Phys. Rev. Lett.* **97**, 083002 (2006).

[17] F. Dolde, H. Fedder, M.W. Doherty, T. Nöbauer, F. Rempp, G. Balasubramanian, T. Wolf, F. Reinhard, L.C.L. Hollenberg, F. Jelezko, J. Wrachtrup, "Electric-field sensing using single diamond spins", *Nat. Phys.* **7**, 459 (2011).
6

# Supplementary Material for

# Beyond ensemble averaging: Parallelized single-shot readout of hole capture in diamond

**Richard Monge[1], Yuki Nakamura[3], Olaf Bach[1], Jason Shao[1], Artur Lozovoi[1,*], Alexander A. Wood[4], Kento Sasaki[3], Kensuke Kobayashi[3], Tom Delord[1,†], and Carlos A. Meriles[1,2,†]**

[1]*Department of Physics, CUNY- The City College of New York, New York, NY 10031, USA.*
[2]*CUNY-Graduate Center, New York, NY 10016, USA.*
[3]*Department of Physics, The University of Tokyo, Bunkyo-ku, Tokyo, 113-0033, Japan.*
[4]*School of Physics, University of Melbourne, Parkville, Victoria 3010, Australia.*

*Present address: *Department of Electrical and Computer Engineering, Princeton University, Princeton, New Jersey 08544, USA.*

†E-mails: tdelord@ccny.cuny.edu, cmeriles@ccny.cuny.edu.

## Contents

## I. Experimental apparatus

Experiments are carried out with a closed-cycle cryostation[1] at a temperature of 9 K (see schematics in Supplementary Fig. 1); a vacuum compatible microscope objective (NA 0.75) focuses the excitation lasers and collects the photoluminescence (PL) from a sample positioned on low temperature nano-positioners[2]. A dichroic mirror on a magnetic mount outside of the cryostat window allows us to steer the emitted PL towards a high sensitivity electron multiplying charge coupled device (EMCCD, Princeton Instrument ProEM-HS:512BX3) or an avalanche photodetector (APD) while maintaining the same excitation path.

Steering mirrors allow us to scan a focused beam across our diamond sample: We use a green laser (532 nm, 1-5 mW) for NV charge initialization and a tunable narrowband laser (637 nm, 0.05-20 μW) to perform photoluminescence excitation (PLE) spectroscopy[2,3] or charge read-out[2,4]. Each laser is controlled by an acousto-optic modulator and a shutter is added to the green laser to prevent any leakage from reaching the sample. A wavemeter (High Finesse) allows us to monitor or stabilize the frequency of the resonant laser[2]; the measured laser powers are attenuated by a factor 3.5 before reaching the back-aperture of the microscope objective. We use a microwave field generated by a 25 μm wire-antenna laid on the diamond to repump the $NV^-$ spin and avoid trapping into dark states under continuous resonant excitation[5].

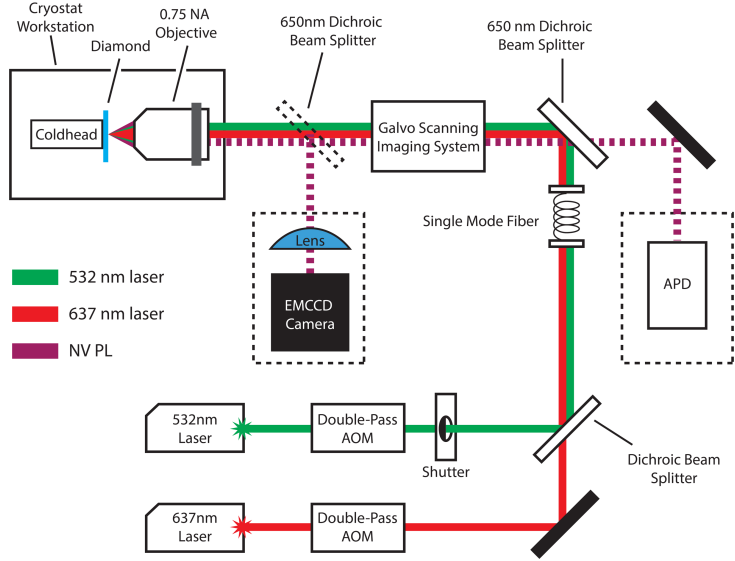

**Supplementary Figure 1 | Schematics of the experimental setup.** Our custom-made microscope can be configured to operate in confocal or widefield mode. The sample sits in a cryo-workstation, and we measure the NV PL with an avalanche photo detector (APD) or an EMCCD camera.

To illuminate a large area of our sample, we exploit the reflection of the focused laser beam on the diamond back-surface: By underfilling the back-aperture of our objective and for a 200 μm thick diamond, this method allows us to generate a 75-150 μm diameter beam at the objective focal plane. We also control the laser incidence angle such that the laser back-reflection is shifted from the laser focal point, which enables us to switch between focused and widefield illumination by changing the beam position with steering mirrors. The collected photoluminescence is filtered with 532-nm notch and 650-nm long-pass filters, and focused on the CCD array with a 25 cm lens yielding a 173× magnification.

All experiments are carried out on a CVD grown electronic grade diamond purchased from Delaware Diamond Knives, which was subsequentially implanted with a 20 MeV $^{14}N$ focused ion beam to create nitrogen doped patterns, and subsequently annealed[6]. Most NV centers we use originate from in-grown nitrogen activated by annealing. Except for the NV density, we did not discern systematic differences with implanted NVs.

## II. Experimental protocols

### II.1 Hyperspectral imaging of NV ensembles

Supplementary Figure 2 illustrates our widefield resonant imaging protocol for capturing PLE spectra across many individual NV centers in parallel. Panel 2a shows both the measurement sequence and the illumination geometry. We begin with green widefield excitation (532 nm), delivered via back-reflection through the diamond substrate, to initialize NV centers into their negative charge state over a large field of view. A tunable red laser (637 nm) is then swept in frequency while illuminating the same region, enabling spectral interrogation of each NV's optical transition.

Panel 2b displays a representative widefield EMCCD image constructed from the integrated PLE signal. Each bright spot corresponds to an individual NV center (or sub-diffraction cluster), with brightness proportional to the aggregate signal collected over multiple frequency sweeps. The image reveals the spatial distribution of NVs, highlighting the capability to monitor many emitters simultaneously.

In Panel 2c, we focus on the site circled in Panel 2b and track the PLE spectra over time. Two distinct NV centers are visible through their characteristic resonances at $\nu_A$ and $\nu_B$. Here, we deliberately apply strong resonant red excitation, leading to controlled photoionization of both $NV^-$ centers after a few frequency sweeps. This results in the disappearance of the optical resonances, as shown in the temporal evolution of the spectral amplitudes (upper right insets). In passing, these observations showcase our ability to track charge state changes at the single-NV level and with sub-diffraction discrimination using resonant spectroscopy.

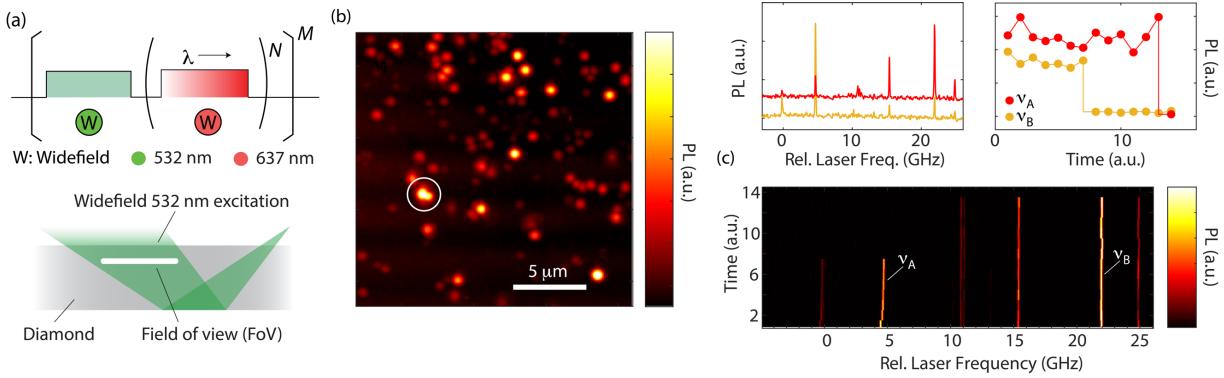

**Supplementary Figure 2 | Hyperspectral NV center imaging.** (a) Measurement protocol and illumination geometry (top and bottom schematics, respectively). We use green widefield excitation from back reflection on the diamond bottom surface to charge initialize NVs over a large area; consecutive frequency sweeps of a tunable red laser acting in the same geometry allows us to record multiple spectra for every NV within the field of view. (b) Example EMCCD image; the brightness of each pixel represents the aggregated intensity derived from the corresponding PLE spectra. (c) PLE spectral series at the site circled in (b). We identify two NVs each featuring characteristic absorption peaks; moderate red illumination intensity leads to selective $NV^-$ ionization, which we capture by monitoring the amplitude of resonances $\nu_a$ and $\nu_b$ as a function of time (upper left insert).

## II.2 Photogeneration of carriers via optical NV center charge-state cycling

NV centers in diamond exist in two stable charge states, negatively charged and neutral, each with unique optical and spin properties. Optical illumination enables reversible charge state conversion between $NV^-$ and $NV^0$, offering both a tool for initialization and a handle for probing charge dynamics in the diamond lattice. Under green light excitation (532 nm), both ionization ($NV^- \rightarrow NV^0$) and recombination ($NV^0 \rightarrow NV^-$) occur via two-step, one-photon processes[7]. For ionization, a single green photon promotes $NV^-$ to its excited triplet state, followed by absorption of a second photon that lifts the excess electron into the conduction band, resulting in $NV^0$. Conversely, recombination proceeds via initial excitation of $NV^0$ into its excited state, followed by absorption of another photon that promotes a valence band electron into the defect level, thus forming $NV^-$. This photo-driven charge cycling not only underpins standard NV initialization protocols, but also makes NV centers active players in the local charge environment.

Beyond their role as emitters, NV centers can also act as localized sinks for photogenerated charge[8-11]. In particular, $NV^-$ centers can be neutralized through the capture of holes diffusing through the lattice, offering a direct optical handle on carrier motion and recombination dynamics at the single-defect level. This sensitivity to ambient charge flow enables NVs to serve not only as passive sensors but as active probes of charge capture processes in complex environments. Supplementary Figure 3 illustrates how to leverage individually-addressed NVs as sources and sinks of charge. Panel 3a outlines the measurement protocol: We begin by initializing all NVs in the field of view into the negative charge state via widefield green illumination. A focused green beam is then used to generate photocarriers at a selected site, while a subsequent red PLE scan probes the charge state of the surrounding NV ensemble. As depicted schematically, photogenerated holes diffuse away from the injection point and are captured by nearby NVs, resulting in local charge neutralization. These charge states are metastable in the dark, with effectively unlimited lifetime, a property previously exploited to demonstrate sub-diffraction optical data storage[12]. Note that, unlike in Supplementary Fig. 2, the red illumination used here can be made sufficiently weak to ensure that the ionization probability remains negligible, allowing for non-destructive readout of the NV charge state[12].

Panel 3b shows PLE images before and after 10 s of focused green illumination at a dense NV cluster (marked by the green circle). We observe a pronounced dimming of $NV^-$ contrast well beyond the illuminated region, indicating that holes travel over micrometer distances before capture. As a control, Panels 3c and 3d repeat the measurement with the green beam placed at a site containing no NVs. In this case, the final PLE image reveals no significant change in $NV^-$ signal, confirming that charge injection is mediated primarily by carrier generation from NV centers. We caution, however, that other defects present in the material — such as vacancy complexes — may also contribute to carrier generation under green illumination, potentially acting as additional charge sources depending on their optical cross-sections and ionization thresholds[11]. We show later that this contribution — especially important in experiments aimed at quantitatively determining the intrinsic hole capture cross section of individual NVs — can be measured and accounted for (see Sections II.7 and II.8 below).

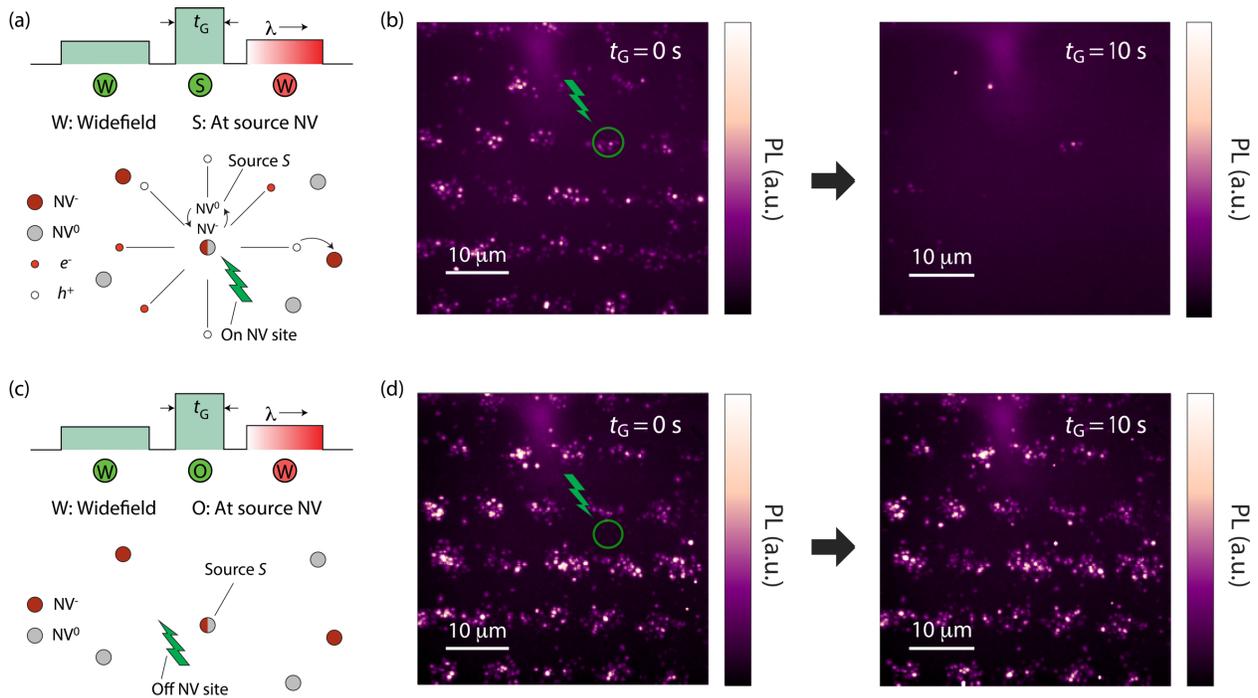

**Supplementary Figure 3 | The NV as a charge emitter.** (a) Measurement protocol and schematics of charge dynamics (top and bottom, respectively). After widefield initialization into NV$^-$, photogenerated holes diffusing from the point of charge injection neutralize surrounding NVs; we measure the resulting charge state of the NV set upon tunable red illumination in widefield geometry. (b) Widefield PLE images of an NV array before and after 10 s of green illumination of a small NV island (green circle in the image); capture of photogenerated holes renders the set dark. (c) Same as in (a) but for green illumination away from an NV site. (d) Same as in (b) but moving the point of focused green illumination to a site hosting no NVs; inspection after $t_G = 10$ s of illumination reveals most NVs remain in the original charge state, thus indicating the absence of photogenerated carriers.

**II.3 Optimized widefield charge readout**

A key challenge in charge readout arises from the fact that the PL intensity at a fixed resonant frequency is sensitive not only to the NV charge state, but also to local electric field fluctuations that can shift the optical transition via the Stark effect. As a result, an NV may appear to ionize simply because its resonance has drifted out of the detection window, leading to potential misidentification. To mitigate this issue, one would ideally sweep the laser across a broad spectral range to ensure full resonance capture; however, such wide scans are time-consuming and limit the number of NVs that can be read out simultaneously.

In Figs. 1 through 3 of the main text, we perform charge readout by sweeping the laser frequency over a comparatively limited frequency window, playing on a tradeoff between time-efficient accumulation of statistics and the number of NVs we are simultaneously reading out. A single charge readout is then performed by acquiring 200 frames at a 10 Hz frame rate while sweeping the resonant laser frequency over a time of 10 s. To ensure addressing the same frequency over prolonged periods of time, the average frequency of the laser is stabilized using a slow enough proportional-integral-derivative (PID) controller. Moreover, the laser frequency is tracked during camera acquisition by calibrating the laser modulation and measuring the frequency before and after a camera acquisition.

Importantly, the fidelity of this charge readout protocol can be compromised by spectral drifts whose amplitude approaches the boundary of the measurement window. Supplementary Fig. 4 presents data for two NVs, one with a mean frequency near the center and the other near the edge of the window. NV$_E$ displays strong systematic frequency drifts associated with an increase of its hole capture rate (see main text Fig. 3). Since its optical resonance frequency remains hundreds of MHz away from the edge, however, we can readily conclude that PL loss corresponds to a change of the charge state and carrier capture. On the other hand, the frequency of another color center, NV$_F$, often drops close to the measurement window boundary; the loss of PL we observe at longer times is therefore more likely correlated with a change of its frequency rather than its charge state, a condition that triggers automated removal from our data analysis see Section II.5).

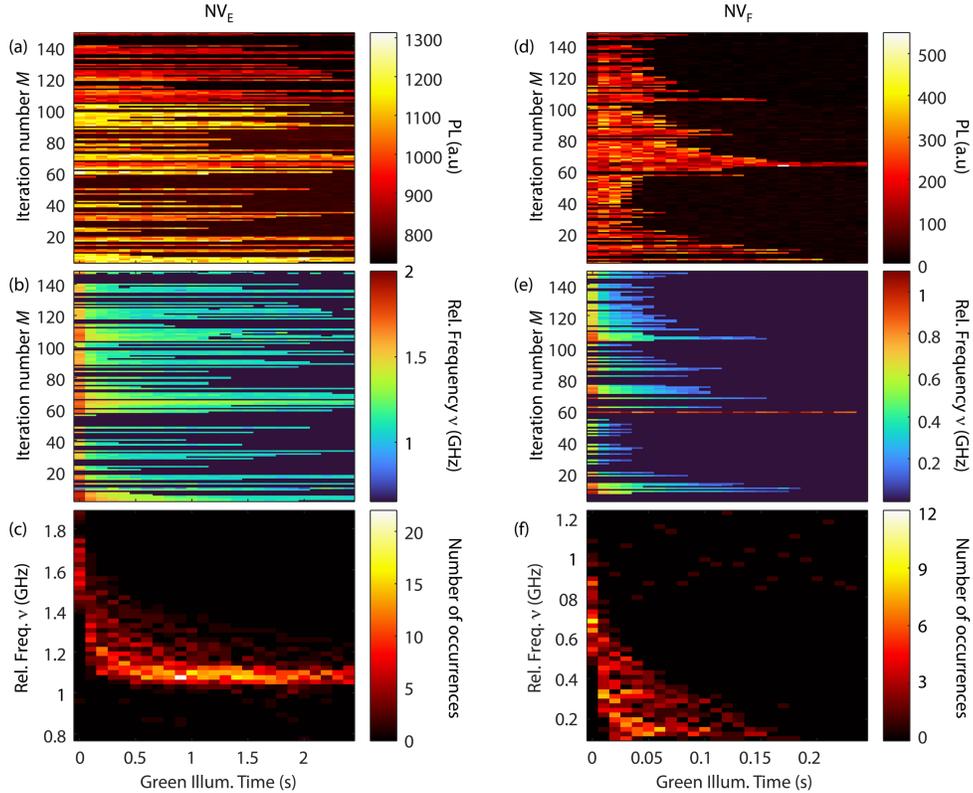

**Supplementary Figure 4 | Charge readout robustness against spectral drifts.** (a, b) Respectively, maximum PL and frequency of the maximum as a function of the green illumination time for multiple experimental runs (iterations) for $NV_E$ (see Fig. 3 main text). (c) Frequency distribution as a function of the green illumination time for all 144 iterations. (d-f) Same as in (a-c) but for another NV in the set, here labeled $NV_F$. In this case, the NV frequency is close to the edge of the measurement window, the maximum PL does not correlate with the charge state but rather with the frequency of the resonance, which shifts with changes of the local charge environment. The laser frequency scan window is 0 to 2 GHz and reference frequency is 470.474 THz.

We determine the read-out–induced NV⁻ ionization rate by measuring charge state transitions in the absence of carrier generation from a nearby source. This yields an average ionization rate of 0.2 Hz, corresponding to an ionization probability of 0.02 per read-out. While this effect can slightly overestimate the capture radius — especially for NVs distant from the source — it does not account for the trend observed in Fig. 2e in the main text.

**II.4 Determining the hole emission rate of source NVs**

Widefield charge state readout makes it possible to multiplex hole capture measurements over hundreds of NVs. This method allows for a substantial speedup compared to the sequential measurement of each NV center (×10 or more depending on readout times and number of NVs). On the other hand, the lower excitation power required for point illumination leads to diminished background fluorescence. Correspondingly, transitioning from a widefield to a confocal geometry (where the readout beam acts locally and we redirect the PL away from the EMCCD camera and into the APD, see above Section I) can accumulate statistics for specific NVs at a faster rate. This method is utilized to characterize the charge cycling rates of NV centers subsequently serving as charge sources (main text, Fig. 1b), and when measuring the capture rates under different charge environment initializations (main text, Fig. 4).

The latter also applies to the parallelized experiments of Fig. 2 in the main text with one caveat: Analysis of the PLE traces stemming from the focused green excitation site reveals the presence of two closely spaced NV centers — denoted $NV_S^{(1)}$ and $NV_S^{(2)}$ — located within the same diffraction-limited spot. To quantify their contribution to the local hole flux, we use the protocol in Fig. 1a of the main text to separately extract their charge conversion dynamics (Supplementary Fig. 5). Automated charge state assignment is performed by applying a fixed PL intensity threshold to the time trace: Signal levels above the threshold are attributed to the bright NV⁻ charge state, while lower values correspond to the dark

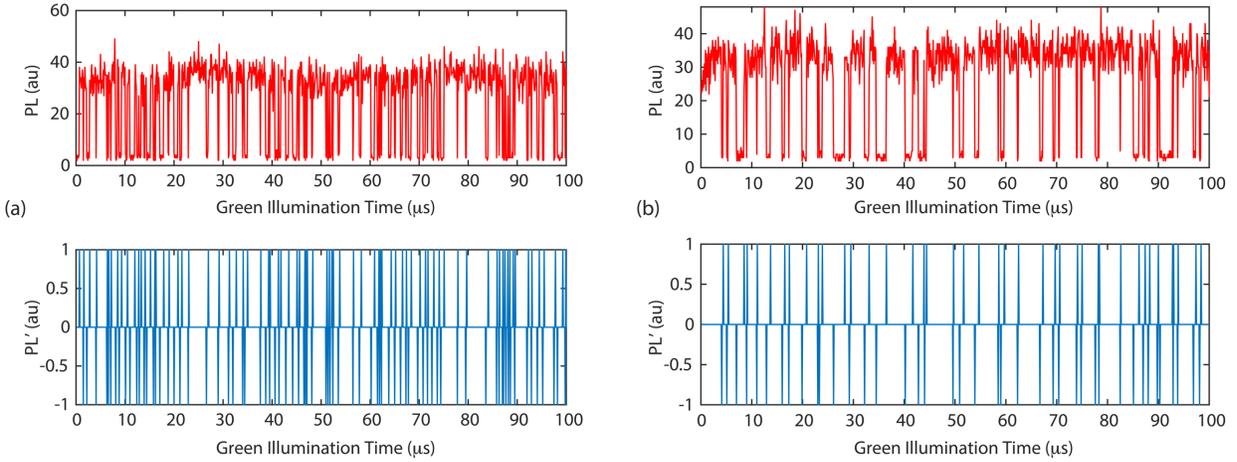

**Supplementary Figure 5 | 2-NV cluster charge cycling rates.** Analysis of the PLE spectra over multiple laser frequency scans indicates the site under focused green illumination in the experiments of Fig. 2 contains not one, but two NVs. (a) PL trace and its derivative after automated charge assignment (top and bottom plots, respectively) for source $NV_S^{(1)}$ in the widefield experiments of Fig. 2 in the main text. Out of 80 cycles, we extract a charge cycling rate of 0.80(9) MHz for an average NV$^-$ fractional population of 0.72(2). (b) Same as in (a) but for the $NV_S^{(2)}$. From 41 cycles, we calculate a rate of 0.41(6) MHz and an average NV$^-$ population of 0.75(2). Throughout these experiments, the focused green beam has a power of 5 mW.

NV$^0$ state. To detect charge conversion events, we compute the time derivative of the PL signal and identify pairs of opposite-sign spikes (a sharp negative followed by a positive, or vice versa) as individual charge cycles.

The charge cycling rate is then defined as the total number of such cycles divided by the integrated green illumination time. Since hole injection occurs during NV$^0$ → NV$^-$ recombination, this rate directly corresponds to the average hole emission rate per NV center. We find that $NV_S^{(1)}$ undergoes 80 such cycles over the measurement interval, corresponding to a recombination rate of 0.80(9) MHz, while $NV_S^{(2)}$ exhibits 41 cycles, yielding a recombination rate of 0.41(6) MHz; the difference likely stems from different excitation efficiencies, in turn, a consequence of distinct NV orientations relative to the green beam linear polarization[12]. We calculate the total hole flux from the cluster as the sum of the two rates, 1.21(11) MHz.

Additionally, we extract the average NV$^-$ population from the same traces by computing the fraction of time the PL signal remains above the threshold. This yields NV$^-$ occupations of 0.72(2) and 0.75(2) for $NV_S^{(1)}$ and $NV_S^{(2)}$, respectively. These values are consistent with those known for bulk NV centers, as extracted from the ratio between the recombination and ionization rates under green illumination[7]. We follow identical protocols to derive the hole emission rates of other source NVs.

**II.5 Data analysis**

We extract the charge states and optical resonances of NV ensembles using the measurement protocols detailed in Section II.1, processing the acquired data through a series of computational steps. As shown in Supplementary Fig. 6a, each dataset consists of a sequence of EMCCD frames (2D spatial arrays) recorded across different laser frequencies, green illumination durations, and measurement iterations. This yields a five-dimensional dataset, where photoluminescence is tracked as a function of position (2D), laser frequency, carrier injection count (i.e., green illumination time), and acquisition index. A single multiplexed run — typically comprising 140 iterations with 25 readouts each — can amount to several hundred gigabytes, consistent with the number of frames and the 16-bit resolution of the EMCCD. The combined use of resonant optical spectroscopy, controlled charge injection, and temporal multiplexing leads to a level of data complexity that significantly exceeds that of multiplexed off-resonant charge readout[13,14] or widefield optical spectroscopy[15].

The main data processing steps are outlined in Supplementary Fig. 6b. Step A addresses thermal drifts that may occur during long acquisitions (typically 6–12 hours for multiplexed measurements). To correct for these drifts, we use the carrier source NV as a spatial reference, recording its position under focused illumination with the EMCCD before each charge initialization cycle; each frame is then shifted accordingly. Given that one EMCCD pixel corresponds to ~90 nm in the image plane, we apply spatial smoothing in Step B using a 2D Gaussian filter with a 250 nm width.

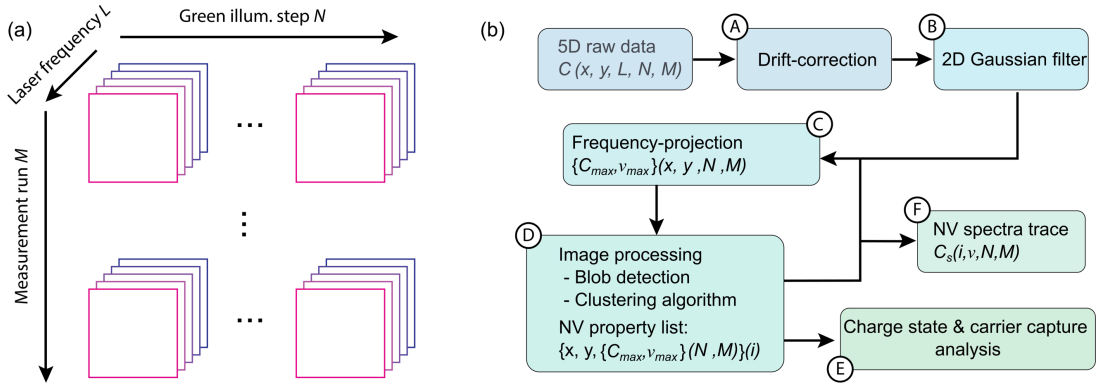

**Supplementary Figure 6 | Data structure and data analysis pipeline.** (a) Our multiplexed measurement takes the form of a 5D dataset: Each square represents a camera frame measuring the photoluminescence ($C_r$) as a function of the position in space. The dataset is constructed according to the protocol previously described by accumulating frames while sweeping the laser frequency (index $L$), green illumination step (index $N$), and repetition index after optical re-initialization (index $M$). (b) Data analysis pipeline: The raw data is sequentially transformed to extract the photoluminescence and optical resonance frequency of a collection of NV centers; $i$ denotes individual NV centers.

In Step C, we reduce the data volume by collapsing the frequency dimension: For each green illumination step $N$ and repetition $M$, we extract the maximum PL across all laser frequencies and record the corresponding frequency value using the known laser modulation calibration. This compresses the dataset by approximately an order of magnitude, yielding two 4D datasets (PL and frequency) from the original 5D measurement.

Step D is the core image processing stage. For each measurement, we identify potential NV centers using a blob detection algorithm, followed by a clustering procedure that groups repeated detections based on spatial proximity. NVs that are not sufficiently resolved or detected consistently across measurements are excluded to streamline the analysis. This results in a final set of NV centers (indexed by $i$, a total of 255 in Fig. 2 of the main text), each characterized by its mean position $(x_i, y_i)$, its maximum PL signal $C_{\max}(N, M)$, and the corresponding peak frequency $\nu_{\max}(N, M)$ as a function of the green illumination step $N$ and repetition index $M$.

The extracted values of $C_{\max}(N, M)$ and $\nu_{\max}(N, M)$ are then used to analyze the charge dynamics of each NV center (Step E). For each iteration $M$, we fit $C_{\max}(N, M)$ with a step function to identify transitions from NV$^-$ (bright) to NV$^0$ (dark). The validity of each fit is assessed based on the magnitude of PL drop relative to the PL noise. Valid fits are then used to build statistics, including NV brightness distributions, which help assign the most probable charge state in ambiguous cases. This analysis yields the charge state of each NV center as a function of both green illumination step $N$ and repetition index $M$, from which we compute carrier capture probability histograms by averaging over $M$, as shown in Figs. 1–3 of the main text. Importantly, $\nu_{\max}(N, M)$ serves to identify cases where PL loss results from the NV optical resonance drifting out of the modulation range. To avoid misclassification, we exclude from the analysis any NV center whose average resonance frequency lies within 250 MHz of the edge of the frequency window.

For most NV centers, the carrier capture probability follows an exponential decay, reflecting a constant capture rate combined with the decreasing NV$^-$ population. In these cases, we extract the capture rate by fitting the decay with a single exponential. When the capture rate is not constant, we instead normalize the capture probability by the instantaneous NV$^-$ population to obtain, for each time step, the conditional probability that an NV$^-$ captures a hole during carrier injection. Assuming this probability remains constant over the time interval, we calculate the corresponding hole capture rate and capture radius at each step. To mitigate noise — especially at later time steps when the NV$^-$ population is low — we average over multiple steps.

In parallel (Step F), we reconstruct spectral traces for each NV center to gain insight into changes in their charge environment. Using the known positions of the NVs, we extract photoluminescence as a function of laser frequency from the 2D spatially filtered data. To improve the signal-to-noise ratio, we apply a 1D Gaussian filter (100 MHz width) along the frequency axis; this yields frequency-resolved spectra for every NV center across the multiplexed dataset, as shown in Fig. 3 of the main text (see also Supplementary Fig. 7 below).

**II.6 Observation of local space charges**

Spatial inhomogeneities of trapped charge—commonly referred to as space charges—are a well-known feature of doped semiconductors[16,17]. Color centers embedded in the semiconductor provide a powerful tool to probe their spatial distribution[17,18], acting as *in situ* local electric field sensors. In our experiments, systematic shifts in the NV optical resonances demonstrate that built-in electric fields can be deterministically manipulated through a combination of optical initialization and carrier injection (see Fig. 3 in the main text and Supplementary Fig. 7). These fields can emanate from the build-up or depletion of space charges, located in the diamond bulk or on the diamond surface[15].

A key advantage of electric sensing using NV PLE is the ability to infer the location and distribution of electric field sources at the nanoscale by employing multiple spatially separated sensors[4,19]. Although a complete determination of the space-charge distribution and amplitude is beyond the scope of this paper, we can draw several conclusions about their nature and impact from our observations and simple electrostatic calculations.

Supplementary Figure 7a shows a PL map from four NV centers. Three of them ($NV_J$, $NV_K$, $NV_L$) are located within 1.2 µm of $NV_D$, which displays strong systematic shifts of its optical resonance (Supplementary Fig. 7e, see also Fig.3a in the main text). The lack of correlation with spectra from the three nearby NVs (Panels 7b through 7d) demonstrates the local nature of the space-charge field. We caution that since NV electric sensing relies on the permanent electric dipole in the excited state manifold, it is subject to a strong orientational dependence[4,19]. However, the fact that none of the three NVs surrounding $NV_D$ senses the same space-charge field strongly suggests that the lack of shifts is not an orientation artifact but rather a reflection of the space charge localization.

To estimate the proximity of the space-charge to $NV_D$, we compare the electric fields $E_{J,K,L}$ stemming from a hypothetical point-charge $\delta_D$ at $NV_D$ and at the other NVs. Using the observed shifts (225 MHz for $NV_D$ and less than 25 MHz for $NV_{J,K,L}$), we estimate that the space-charge region is less than 600 nm away from $NV_D$. In addition, discretization of certain shifts indicates single charge traps are between 50 and 100 nm away from $NV_D$, which might form the tail of an extended space-charge region. Depending on its exact proximity, the overall space-charge is composed of 5 to 70 elementary charges (Supplementary Fig. 7f).

These observations rule out the role of the diamond surface (10 µm away in our experiments) as the source of space-charges, and point at regions containing high density of charge traps, which get ionized during optical illumination and neutralized under carrier injection. While a definitive identification is not yet possible, these observations are consistent with these traps being substitutional nitrogen centers $N_S$, which are present at 0.1-1 ppb in our samples (18-180 defects per cubic microns) and can be photo-ionized under green illumination.

Local space charges and their associated electric fields can strongly perturb carrier capture by NV- centers, even in the absence of direct ionization. For instance, a configuration of just seven positive charges located 300 nm from an NV- can induce optical Stark shifts exceeding 200 MHz. In optimal geometries (e.g., charges situated opposite the capture path), these fields overcome the Coulomb potential of the NV- center as close as 180 nm from the defect, significantly suppressing carrier attraction. While intrinsic capture radii can exceed 300 nm at low temperatures, such external perturbations may reduce this range by more than half. This is consistent with our measurements: For NVs 10 µm away or more from the source, 40% of those with capture radii below 30 nm exhibit systematic unidirectional shifts in optical resonance larger than 250 MHz. In contrast, only 5% of NVs with capture radii above 50 nm display such behavior.

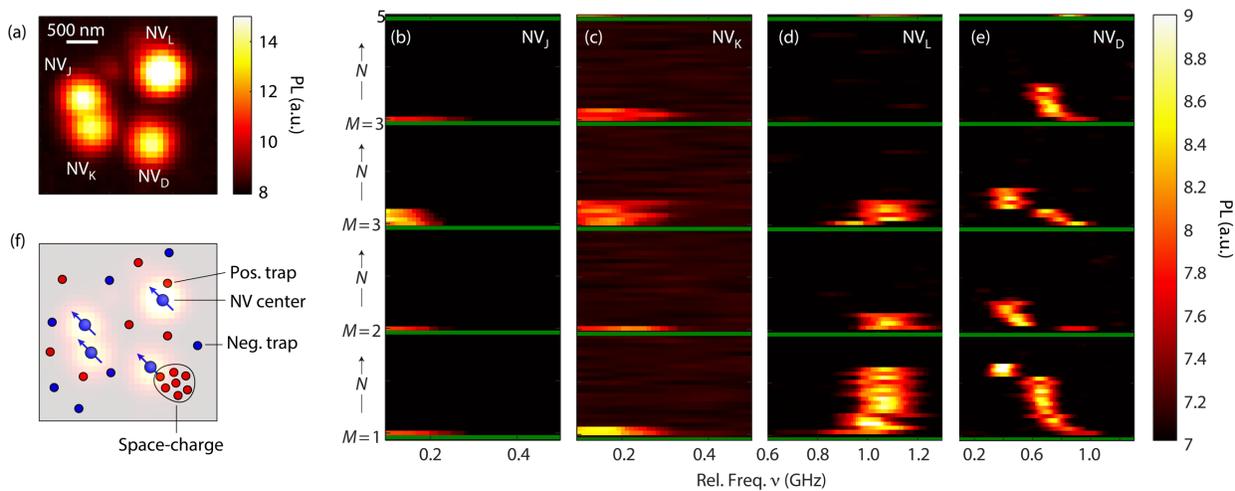

**Supplementary Figure 7 | Localization of space-charges.** (a) PL map of four nearby NV centers 12 µm away from a charge source. (b-e) Spectra of the same NVs upon application of the protocol in Fig. 1a of the main text; as in Fig. 3, horizontal green lines indicate successive widefield initialization with green light. (f) Example of a space charge configuration leading to the measured spectra. The faint background reproduces the image in (a) as a reference. Throughout these experiments, the laser reference frequency is 470.474 THz.

## II.7 Capture of photogenerated holes in neutralized environments

The hole capture cross section measured in experiments typically reflects an effective value, shaped not only by the intrinsic probability of carrier capture at the NV center but also by external factors such as excess carrier energy and the presence of background charged traps. These traps — activated during widefield green illumination — can either screen the Coulomb potential of the NV or act as competing capture sites, both leading to a suppression of the apparent hole capture probability.

To isolate the intrinsic capture cross section of the NV — the quantity governing charge capture in an ideal, unscreened and thermalized environment — we introduce a multi-step protocol (Supplementary Fig. 8) that corrects for these distortions. First, we thermalize the injected carriers by using probe NVs at varying distances from the charge source (respectively, $NV_P^{(j)}$, $j = 1,2,3$ and $NV_S$ in the image of Panel 8a) allowing hot holes to cool down to the lattice temperature before encountering the NV (we discuss this effect separately in Section III below). Second, we precondition the local environment by neutralizing residual charged traps prior to measurement. This is achieved through remote carrier injection from a separate NV cluster whose photoionized carriers are captured by traps throughout the region. While this strategy is not strictly necessary to observe hole capture, it significantly reduces screening and competitive capture effects, enabling us to better approach the intrinsic NV response.

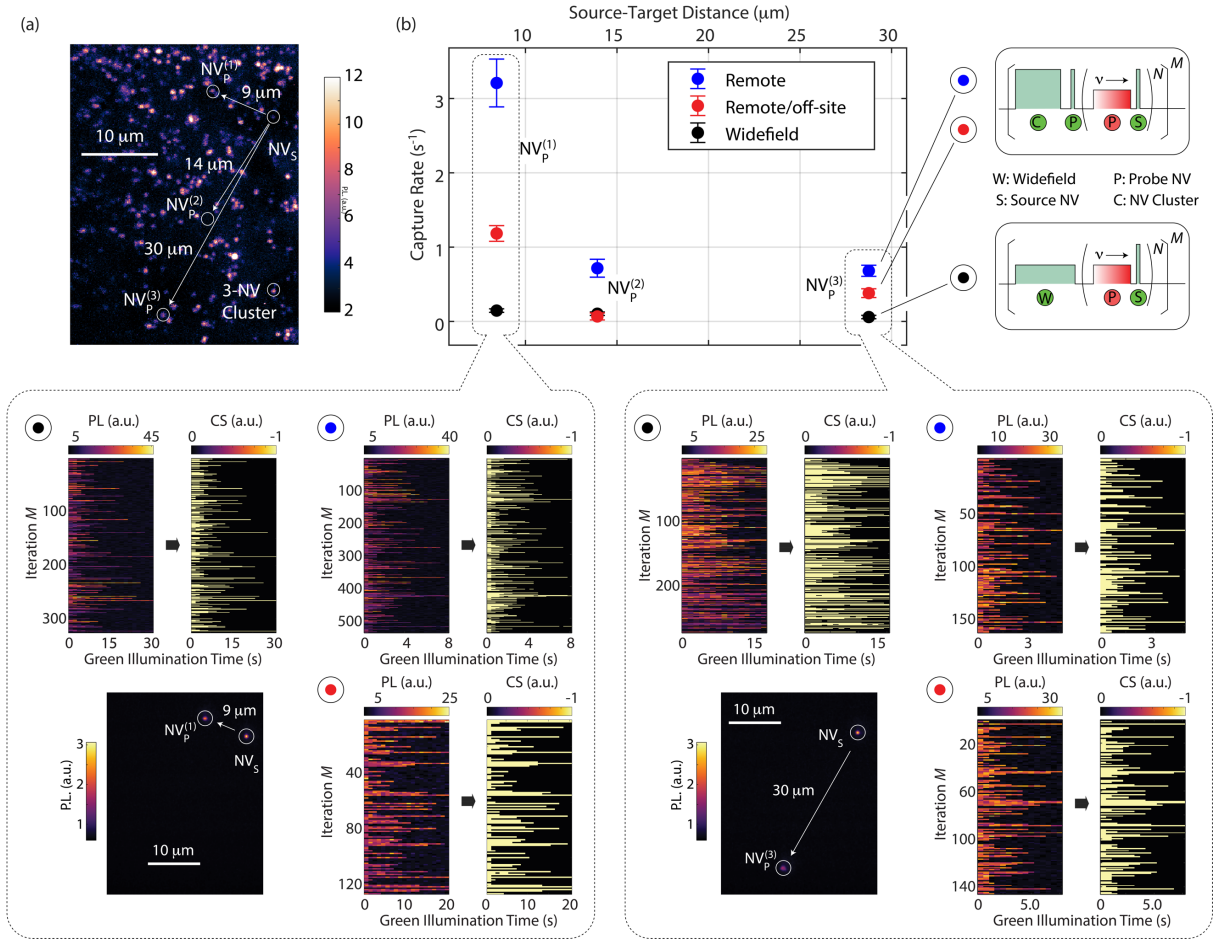

**Supplementary Figure 8 | Charge initialization via remote carrier injection.** (a) Confocal image of an area of the crystal with a collection of NVs. We examine the capture of photo-generated carriers generated by select probe NV centers in the set at varying distances from the source, $NV_S$. Note that not all NVs are visible in this image; this is the case of $NV_P^{(2)}$. Also circled is the 3-NV cluster (bottom right) serving as an alternative charge source during initial neutralization. (b) (Right insets) Measurement protocols corresponding to widefield charge initialization (bottom) and trap neutralization from a remote source followed by focused probe NV activation (top). (Main) Measured hole capture rates for three different probe NVs (corresponding to those shown in Fig. 4e of the main text) as derived from three different protocols. The "Remote/off-site" protocol is identical to the "Remote" protocol except that we move the beam away from the source NV during the $N$-loop. The presence of non-zero capture rates points to other background sources besides $NV_S$. (Bottom inserts) PL time traces for probe $NV_P^{(1)}$ and $NV_P^{(3)}$ as measured (left) and after automated charge state assignment (right) for each protocol in use. Shown in the bottom left corner is the widefield image under red illumination following remote neutralization; notice that only the source and probe NVs are visible.

Supplementary Figure 8b shows the hole capture rates for three NV centers as a function of the distance to the carrier source and for different initialization and hole injection conditions: Generally, neutralized conditions lead to much greater carrier capture rate. To ensure that we attribute hole capture to carriers originating only from the designated source NV, we implement a background correction step. Specifically, we repeat the measurement sequence with the green laser beam displaced from $NV_S$ to an adjacent empty site (Supplementary Fig. 8b, "Remote/off-site" protocol). This allows us to quantify any residual hole injection that might occur from other photoactive defects or background processes (see Section II.8 below).

The inset panels in the lower half of Supplementary Fig. 8b illustrate this protocol for two probe NV centers, $NV_P^{(1)}$ and $NV_P^{(3)}$, including raw and processed PL time traces; the main plot in the upper half displays the corresponding extracted hole capture rates across different initialization schemes. Throughout these experiments, the readout laser power ranges from 50 to 200 nW, the frequency span during a sweep is 900 MHz, and the sweep rate is 2 Hz; charge readout of a probe NV relies on twenty to thirty 20-ms-long frames using the EMCCD. To determine the charge cycling rate of $NV_S$, we intercalate readout intervals with 400 ns pulses of green light.

The data confirms that environmental preconditioning and background subtraction both contribute to a more accurate estimation of the hole capture cross section. Under optimized conditions, we observe effective capture radii approaching 200 nm, which is close to the Onsager radius at 9 K (∼300 nm) — the scale at which Coulomb attraction can overcome thermal motion to bind a mobile carrier. This agreement supports the interpretation that, once corrected for external effects, the NV hole capture process proceeds via long-range, Coulomb-assisted mechanisms, with an intrinsic cross section fundamentally set by equilibrium thermodynamic constraints.

**II.8 Hole generation by neutral dark traps**

Surprisingly, we find that the neutralization step enables hole injections when illuminating nearly any spots not containing NV centers. Completely absent without the neutralization step, this background hole capture rate depends on the chosen illumination sites, and on the probe NV position. This contrasts with the results in Supplementary Fig. 3 and with previous observations[11] where single dark defects were spatially mapped via their hole injection at room temperature in the same sample. The lack of correlation of the background hole capture rate with distance to the source (see Supplementary Fig. 8b) points at a role of the target NV environment, possibly through the back-reflection illumination.

To further investigate this phenomenon, we attempt to charge-initialize an NV center via back-reflection illumination after prolonged neutralization of the environment. Supplementary Figure 9a shows the PL measured as the green illumination time is increased: Interestingly, we observe a change in behavior over several minutes. Over longer timescales, we measure a charge cycling rate of 17 mHz and an $NV^-$ population above 70%, which is consistent with measurement performed under widefield initialization. Over shorter timescale, however, the $NV^-$ charge state remains mostly neutral, which points to hole capture as the dominating charge dynamics (the case in Supplementary Fig. 8b). Our observations indicate that a reservoir of photo-ionizable charge traps builds up upon charge initialization of the environment to neutral, which can then generate holes across our sample once illuminated, and until the reservoir is depleted. The traps constituting that reservoir must be predominantly acceptors ionizing (but not recombining) under 532 nm illumination, possibly substitutional boron, vacancies, or surface acceptor states[20].

Supplementary Figures 9b through 9d give a qualitative description of the state of these charge traps under different conditions. Under focused illumination of a source NV (i.e., the neutralization protocol, Panel 9b), the trap state is set by the competition between photo-ionization and carrier capture. Based on our measurement of electric field fluctuations, we can assert that outside of the focused laser beam and close enough to the carrier source, charge traps are maintained at a low occupancy (less than 0.5 %, see Section III.1). In regions with stronger illumination (e.g., directly above and below the laser focus spot), these populations likely increase. Once the laser focus spot is moved away from an NV center (Supplementary Fig. 9c), previously neutralized traps become available for photo-ionization, leading to a flux of electrons and hoes. While stronger around the laser spot, photo-ionization will also occur over the large area illuminated by the diamond back-reflection: Though single defect rates are likely low, the amount of carriers injected will remain substantial due to the number of defects affected, the single photon-nature of the process, and the ability of carriers to be transported over tens of micrometers. Finally, under prolonged widefield illumination (∼10 min), charge traps are mostly charged, and hole generation and/or transport is suppressed. Note that as in Ref. [21], we name "charged" an environment that may have an overall neutral charge but is constituted of ionized impurities of both charges.

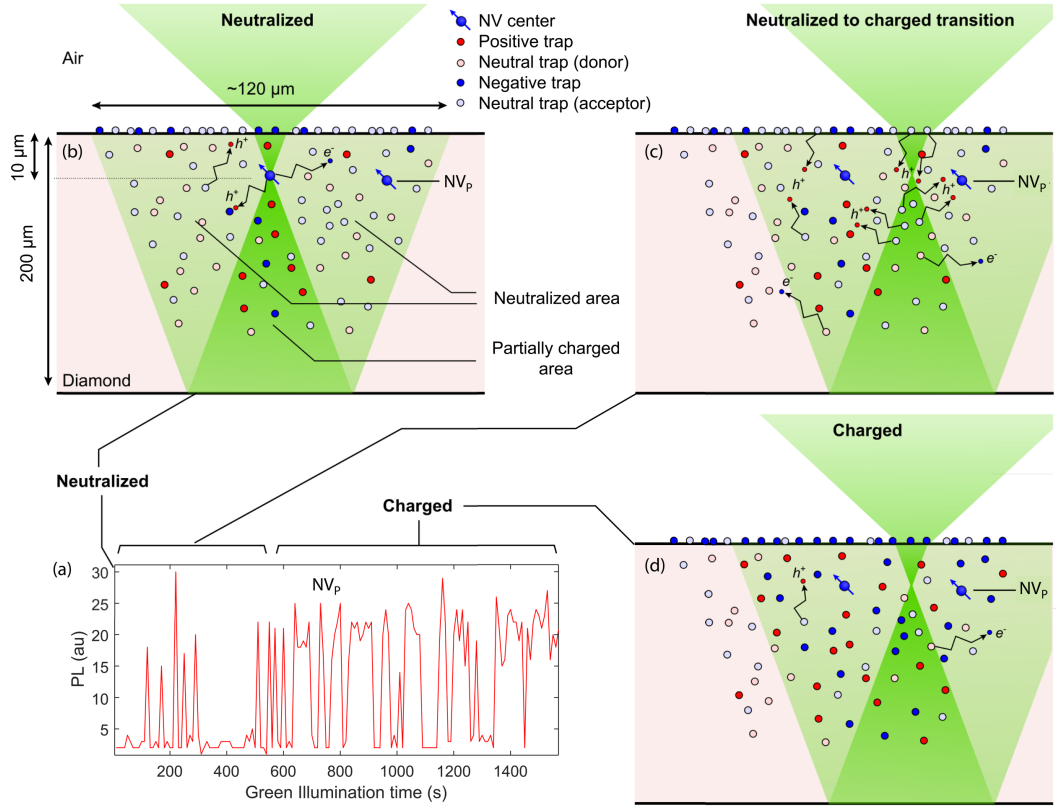

**Supplementary Figure 9 | Hole generation during dark traps charging.** (a) Charge state of a probe NV over prolonged periods of time after neutralization of the environment; here, focused green illumination is carried out continuously at a site with no source NV. Acceptor photo-ionization likely causes charge instability until traps are charged and cannot be photo-ionized. (b) Neutralization via carrier injection at a source NV. Traps are mostly neutral due to faster carrier capture than photo-ionization. (c) Charging: photo-ionization of neutral traps following displacement of the beam and absence of carrier injection at an NV injects carriers. (d) Traps are mostly charged and cannot be photo-ionized. For simplicity, traps outside of the illumination areas are omitted.

### III. Modeling

#### III.1 Neutralization measured via spectral diffusion

Fluctuations of optical resonances (spectral diffusion) directly inform us on the surrounding charge environment. While systematic drifts and discrete shifts can be used to localize specific field sources (space charge or single dark traps), stochastic fluctuations — even unresolved — can also be used to shed light on the average charge environment. Reduced fluctuations — as observed upon neutralization of the charge environment (Fig. 4a in the main text) — are symptomatic of a decrease in charge densities. To quantify this effect, we perform Monte Carlo simulations of the charge environment and its impact on NV center optical resonances.

Specifically, we look at the relationship between the trap density $\rho_t$, the average occupancy $\psi$, and the spectral fluctuations, which we quantify using the standard deviation $\kappa$ of the optical resonance. For a given trap density and average occupancy, we generate random positions for charge traps around an NV center, populate them according to their average occupancy, and calculate the overall field at the NV center. The field is then translated to a frequency shift using the NV electric dipole[22] along the center's symmetry axis. For each simulation run, we iterate over 500 charge occupancy configurations per trap position to calculate the variance and iterate 500 sets of trap positions to average it. The defect density is swept between 3 and $10^4$ defects per cubic micron and the average trap occupation between $10^{-3}$ and 0.5. For each charge density, we vary the simulated volume to have a minimum of 10 charged traps around the NV centers; we disregard the sign of the charge since traps are positioned randomly. For simplicity, we also make the approximation of a single type of trap with the same average occupancy.

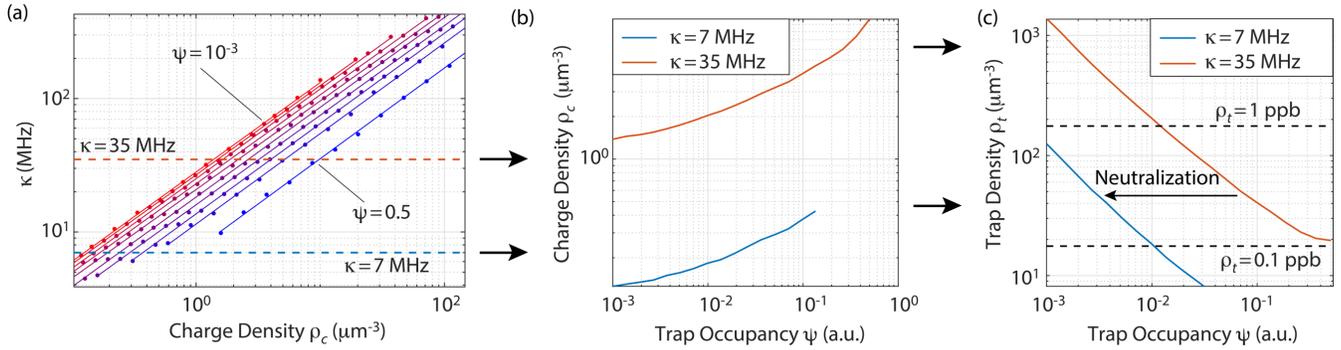

**Supplementary Figure 10 | Monte Carlo simulations of spectral diffusion.** (a) Average standard deviation of the optical resonance as a function of the charge density for different charge trap population. (b) Charge density required to yield 35 and 7 MHz standard deviations as a function of the charge trap occupancy. (c) Same as in (b) but for trap (defect) density instead of charge density.

Supplementary Figure 10 shows the simulation results: In Panel 10a we plot the average value of the standard deviation as a function of the charge density $\rho_c = \rho_t \cdot \psi$ for different trap occupancies. We find that a lower trap occupancy leads to a stronger spectral diffusion, such that different charge densities can lead to the same amount of spectral diffusion (Supplementary Figs. 10a and 10b). Panel 10c shows the trap density leading to specific standard deviations as a function of the trap occupancy. For the measurements of charged and neutralized environments in Fig. 4 of the main text (respectively featuring standard deviations of 35 and 7 MHz), we find possible trap densities ranging from 10 to 1000 charges per cubic micron (0.06 to 6 ppb), in line with expected densities of defects found in electronic grade diamonds (0.1 to 1 ppb of nitrogen or boron). Note that these numbers are also consistent with previous measurement of trap occupancies using NV clusters[4]: using the measured occupancies between 0.01 to 0.16 yields trap densities $\rho_t$ ranging from 0.2 to 0.9 ppb. Finally, for a constant defect density of 0.5 ppb, we find that neutralization corresponds to a 25-fold reduction of the trap occupancy, moving the charge density $\rho_c$ from 2.4-10 to 0.1-0.2 charges per cubic micron (i.e., from 14-57 ppt to 0.6-1.1 ppt).

Note that in all our estimations of the impact of charge densities and trap occupancies on spectral diffusion, cases with complementary occupancies (e.g., 0.95 and 0.05 occupancies) cannot be distinguished. Under charged conditions, it is possible that the occupancy is 0.95 to 0.9 rather than 0.05 to 0.1, which would imply a much stronger reduction in the charge density. This degeneracy can be lifted by combining trap neutralization with the use of NV clusters to identify the charge polarity via charge trap triangulation[4].

### III.2 Diffusion and capture modeling

Modelling carrier dynamics in semiconductors[23,24] typically employs two distinct approaches: The first method relies on ensemble averaging, considering large carrier concentrations and solving Boltzmann transport equations containing explicit diffusion and drift terms[25]. In the present work, individual carriers are sequentially generated and captured, which motivates an explicit treatment of each carrier's trajectory and phonon scattering events through Monte Carlo statistical simulations. This approach was employed in Ref. [26], introducing impurities at deterministic locations that can act as scattering or trapping centers, and modeling carrier via cascade phonon emission[27], using a cutoff in total carrier energy. By incorporating elastic and inelastic acoustic and optical phonon scattering as well as the attractive Coulombic electrostatic of the charged defect, this approach efficiently modeled the giant capture cross section observed in diamond under ambient conditions[6].

At low temperatures, the reduction of the carrier kinetic energy predicts a dramatic enhancement of the capture cross section, with expected capture radii up to ~300 nm at 10 K, for thermalized carriers. In addition, the suppression of acoustic phonon scattering increases mobility, leading to hole mean free paths between 4 and 25 μm[28]. However, the combination of large capture radii and long mean free paths gives rise to other limiting mechanisms, which are typically curtailed by larger phonon scattering rates under ambient conditions.

Here, we use a simplified phonon-scattering and carrier capture model but integrate a broader range of effects such as carrier thermalization and coexisting ionized impurities (see Section II.8). In particular, we rely on the capture radius modeled in Ref. [27] for varying temperature, noting that for carrier temperature above the lattice temperature (9 K), the simulated capture radius is overestimated: While Ref. [27] takes into account the carrier kinetic energy, it assumes the crystal lattice is at the same temperature and therefore overestimates phonon scattering - which enables carrier relaxation and capture. Our aim is to (*i*) demonstrate that our model and understanding is in qualitative agreement with our observations, (*ii*) draw conclusions regarding the initial carrier temperature, and (*iii*) provide a framework for more advanced and accurate simulations.

In the future, quantitative modeling would enable NV arrays to measure quantities such as ionization thresholds, phonon scattering rates, and scattering cross-sections of ionized impurities. Contrary to typical average measurements, the use of collections of individually addressable point defects allows for controlling and monitoring ionized impurities and for the simultaneous observation of carrier capture at various distances (and temperatures) from the injection point.

### III.3 Modeling carrier thermalization

In our experiments, we generate holes via photoexcitation of an $NV^0$ using a 532 nm (2.33 eV) laser. The process involves the absorption of two photons[7]: The first one excites $NV^0$ into a higher energy orbital and the second excites a valence electron, transforming the color center into $NV^-$ while generating a hole. The exact nature of the process is the subject of active research, and determines the energy of the photo-generated hole. In particular, the energy of the orbital from which the second electron is excited changes the ionization threshold and therefore the excess energy brought by the second photon. For example, in Ref. [29], ionization from the $^2A_2$ excited state is considered, which "leaves a hole deep in the valence band at around 1.2 [or] 1.6 eV from the valance band maximum". However, more recently Ref. [30] suggested ionization could occur from the $^4A_2$ shelving state with a much higher ionization threshold (2.3 eV), leaving only 30 meV of excess energy.

In practice, any energy above the lowest optical phonon mode energy (160 meV) is quickly dissipated through optical phonon scattering: The short scattering time (below 10 fs[31]) only lets 1 eV carriers travel a few tens of nanometers before losing the excess energy above optical phonon modes. Once below 160 meV, scattering with acoustic phonons dominates the thermalization and occurs on much longer time scales: Mobility measurements at 13 K have shown relaxation times due to acoustic phonon scattering of 146 ps and 537 ps, which corresponds to mean free paths of 4.3 µm and 25 µm, for heavy and light holes respectively. We extrapolate these scattering rates to an arbitrary carrier temperature by using the formula for the acoustic-phonon deformation-potential in a 3D parabolic band[21]. Given the $\sqrt{E}$ dependence, where $E$ is the carrier kinetic energy, we expect a constant mean-free path, with the higher velocity compensating a higher scattering rate. Note that these values are in reasonable agreement with the relaxation of hot carriers measured via pump-probe experiments[32-34].

We numerically simulate the diffusion and thermalization of carriers via a Monte Carlo method. Specifically, we iteratively compute the random walk of carriers of a given kinetic energy initialized at a point source with a random momentum orientation. For simplicity and due to their stronger representation, we only consider heavy holes with an effective mass $m_{hh}^* = 0.663\ m_0$ [35] and scattering rates extrapolated from Ref. [26] as explained above. At each scattering event, we reduce the carrier energy by $k_B T$ where $T = 9$ K and randomize the momentum orientation. Once the carrier has thermalized to 9 K, the kinetic energy is kept constant, but we continue to randomize the momentum through scattering until the carrier is captured. We note that in our simplified model, each scattering event that fully randomizes the momentum physically consists in scattering by multiple phonons (both absorption and emission and at various energies), each imparting a small change to the carrier momentum. A more accurate model for carrier scattering under similar conditions, as described in Ref. [27], would consider both absorption and emission of phonons at all available energies. On a related note, we model the large density of charge trap at the diamond surface (10 μm above the NVs) by considering that it reflects and absorbs half the carriers reaching it.

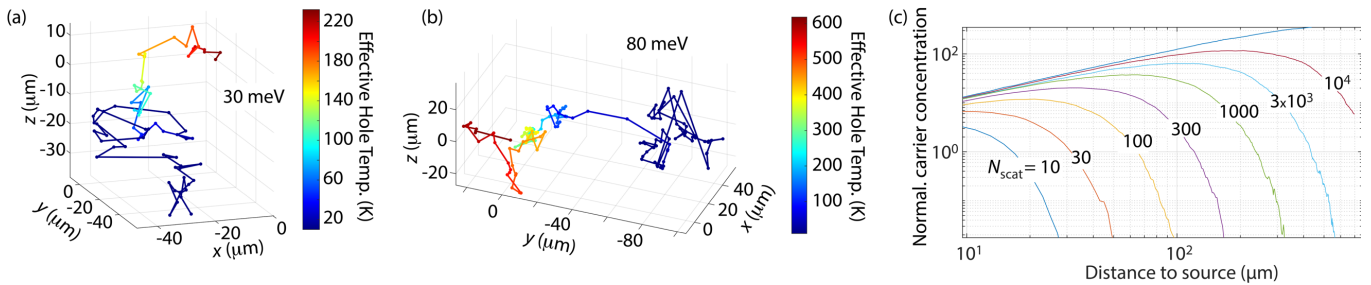

**Supplementary Figure 11 | The impact of hole temperature and lifetime.** (a) Hole trajectory as derived from a Monte Carlo run for a 30-meV hole. (b) Same as in (a) but assuming the starting hole energy is 80 meV; notice the longer thermalization distance. (c) Carrier concentrations integrated over a carrier lifetime and averaged over many trajectories for different carrier lifetimes (counted in scattering events). Concentrations are normalized by those corresponding to ballistic trajectories. Throughout these simulations, carrier momenta are randomized every 4.3 µm.

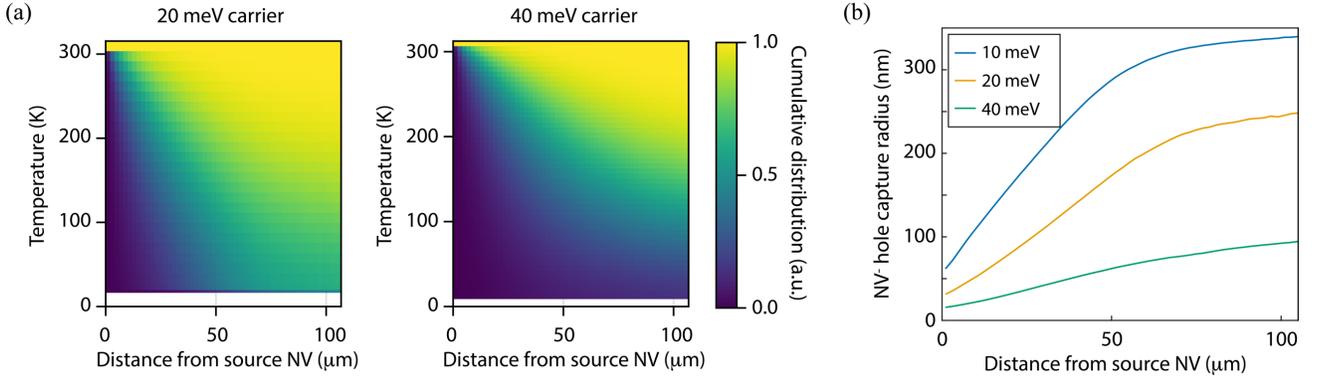

**Supplementary Figure 12 | Impact of the carrier initial energy** (a) Cumulative distribution function of carriers as a function of temperature for varying distance from the source assuming the starting hole energy is 20 and 40 meV (left and right plots, respectively). Notice the longer thermalization distance for more energetic carriers. (b) Calculated effective NV$^-$ hole capture radius vs distance to the carrier source as derived from Monte Carlo simulations for different initial hole energy conditions; negatively charged NVs exposed to cooler holes exhibit greater capture radii.

Supplementary Figures 11a and 11b show the trajectory and temperature of carriers emitted by the same NV for different initial kinetic energies in the absence of capture: As expected, higher energy carriers propagate over a much longer distance before thermalizing. As the carriers undergo random walks, they can also double back and return close to the point of origin after thermalization, thereby enhancing the capture probability. To quantify this effect, we evaluate the carrier concentration integrated over the carrier lifetime: The position of each scattering event is recorded and their density evaluated. In Panel 11c, we illustrate the impact of carriers lingering for prolonged time by showing the result of a simplified simulation: We consider thermalized carriers undergoing momentum randomization every 4.3 μm and vary the carrier lifetimes (counted in the number of scattering events). We normalized the carrier concentration (number of scattering events per volume) by the concentration of carriers under ballistic propagation. We find that for short enough distances, a random walk leads to a substantial increase in the carrier concentration, especially when the carrier lifetime is substantial. This highlights the need to correctly model the carrier loss, here dominated by charge trap capture. Similarly, a more accurate modeling of phonon scattering will change the momentum relaxation rate and impact the calculated capture rate.

To evaluate carrier capture, we take a charge density of $\rho_c = 0.12$ μm$^{-3}$ (0.71 ppt), as estimated in Section III.1, and apply a capture rate per traveled distance of $R_L(T_c) = \pi r_c^2 \rho_c$, where we use Ref. [27] to extract the capture radius, $r_c = r_c(T_c)$, as a function of the carrier temperature, $T_c$. We then iteratively simulate the diffusion of many carriers and record the position and temperature of capture and scattering events to build statistics. Supplementary Figures 12a and 12b show the cumulated distribution of carrier temperature (vertical axis) as a function of the distance to the source for two different initial carrier energies. We clearly observe a change in the temperature distribution with distance to the source: Carriers are hotter at shorter distances, which explains the change in capture radius. This transition is much slower for a carrier of higher energy. In Supplementary Fig. 12c, we confirm the impact of this effect on the effective carrier radii, averaged over carriers of all energies: The capture radii grow with distance to the source and saturate at a shorter distance for cooler carriers. Given our observations (Fig. 4 in the main text and Section II.7 above), our modeling favors the case where photogenerated carriers have a lower energy of 10-20 meV, which points to photo-ionization from the shelving state (20 meV calculated from ab initio).

**III.4 Role of screening by compensating charges**

Compensating charges tend to inhibit carrier capture by screening the field of ionized impurities[36]. Here, we study this effect with a toy model considering the classical trajectory of a carrier approaching a complex of two charge traps of opposite signs. For each relative position of the charge traps, we calculate the trajectories of carriers originating from random positions under the charge trap with constant upward velocity (Supplementary Fig. 13a). We consider the carrier captured if the Coulomb energy exceeds the carrier initial energy, and record the effective capture cross-section as the surface corresponding to initial conditions for which carriers are captured. Supplementary Fig. 13b shows the evolution of the capture radius with respect to a reference point as a function of the distance to the compensating charge and the carrier initial temperature. Here, we average over random orientation of the trap complex, since most carriers are deflected several times before reaching their target. While this model neglects phonon scattering — key to enable energy dissipation and carrier capture — it gives a qualitative account of the expected effect. In particular, we observe that screening gains importance for lower kinetic energy, and substantially reduces the capture radius for distance to the compensating charge in the order of a few hundred nanometers.

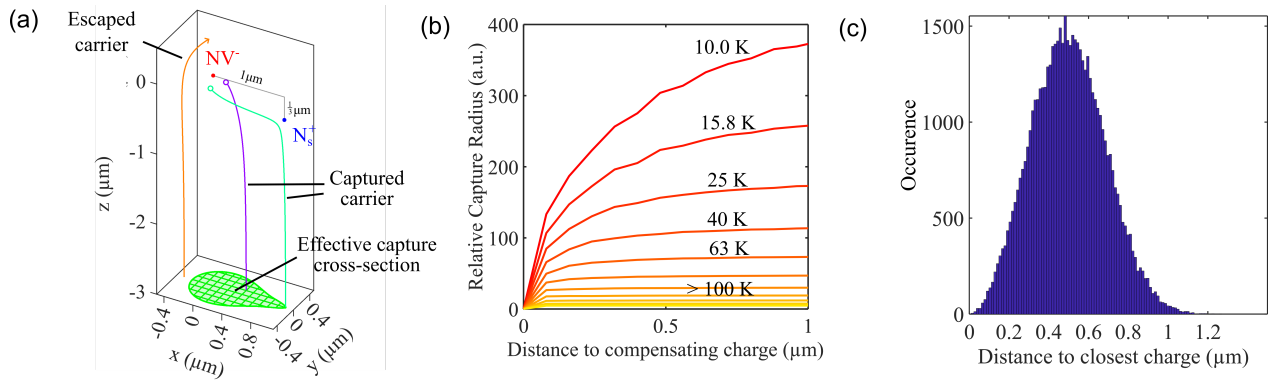

**Supplementary Figure 13 | Role of screening by compensating charges** (a) Classical trajectories for thermalized carriers approaching a target trap (NV⁻) in the presence of a compensating charge ($N_s^+$). (b) Calculated relative effective capture radius as a function of the distance to the compensating charge for varying carrier temperature. (c) Distribution of distances for the closest compensating charges for 150,000 charges with a 0.03 ppb concentration.

This picture is consistent with our estimation of the charge density: Supplementary Fig. 13c shows the distribution of distances to the closest compensating charge for a 0.03 ppb concentration of charges, as previously estimated from PLE for the charged initialization. We conclude that most NV centers are expected to be affected by nearby compensating charges, with the full complexity of the charge landscape — such as discussed in Section II.6 — expectedly leading to stronger effects.